\documentclass[aps,prd,twocolumn,superscriptaddress]{revtex4}

\usepackage{graphicx}
\usepackage{dcolumn}
\usepackage{eufrak}

\renewcommand{\vec}[1]{\mathbf{#1}}
\def\({\left(}
\def\){\right)}
\def\[{\left[}
\def\]{\right]}
\def\be{\begin{equation}}
\def\ee{\end{equation}}
\def\bea{\begin{eqnarray}}
\def\eea{\end{eqnarray}}
\def\ba{\begin{array}}
\def\ea{\end{array}}

\def\ga{\gamma}

\def\de{\delta}

\def\De{\Delta}

\def\ep{\epsilon}

\def\om{\omega}

\def\0{\mathbf{0}}

\begin{document}


\title{Microlensing modulation by quadrupole variation}

\author{Florian Dubath}
\email{florian.dubath@physics.unige.ch}
\affiliation{D\'epartement de Physique Th\'eorique, Universit\'e de Gen\`eve,
24 quai Ernest-Ansermet, CH-1211 Gen\`eve 4\\}

\author{ Maria Alice Gasparini}
\email{alice.gasparini@physics.unige.ch}
\affiliation{D\'epartement de Physique Th\'eorique, Universit\'e de Gen\`eve,
24 quai Ernest-Ansermet, CH-1211 Gen\`eve 4\\}

\author{ Ruth Durrer}
\email{ruth.durrer@physics.unige.ch} \affiliation{D\'epartement de
Physique Th\'eorique, Universit\'e de Gen\`eve, 24 quai
Ernest-Ansermet, CH-1211 Gen\`eve 4\\}

\date{\today}

\begin{abstract}
We investigate microlensing in the case where the lens is considered
as an extended
object. We use a multipolar expansion of the lens potential and show
that the time-varying nature of the quadrupole contribution allows to
separate it from the mass and spin contributions and  leads to
specific modulations of the amplification signal. As example we 
study the case of binary system lenses in our galaxy. The
modulation is observable if the rotation period of the system is
smaller than the time over which the amplification is significant
and if the impact parameter of the passing light ray is
sufficiently close to the Einstein radius so that the
amplification is large. Observations of this modulation can reveal
important information on the quadrupole and thus on  the
gravitational radiation emitted by the binary lens. Even if not 
observed directly, because of their importance the quadrupole
modulation has to be included in the
error budget for high magnification ($\mu\leq7$) microlensing events. 

\end{abstract}


\maketitle

\section{introduction}

The importance of the quadrupole of a binary system relies mainly
in its connection to gravitational radiation via Einstein's famous
quadrupole formula~\cite{Einstein}. This formula is beautifully
confirmed by Taylor's binary pulsar~\cite{taylor,will}, the
indirect proof of the existence of gravitational waves, for which
Hulse and Taylor have been awarded with the Nobel price in 1993.
Direct detection of gravity waves will (hopefully) be realized in
the next few years by the numerous experiments operating today.

The question whether gravitational waves from binary systems can
be detected via their effects on the propagation of photons has
been addressed repeatedly in the past. In \cite{labeyrie} to
\cite{kope} the deflection angle and
the time delay caused by a gravity wave passing through the path
of the photon is analyzed from different points of view, and only in
1998 the solid conclusion to this problem was given by Kopeikin et
al.~\cite{gwinn}: these effects are too faint to be detectable today.

\begin{figure}
 \begin{center}\ \vspace{2mm}\\
\includegraphics[width=9cm]{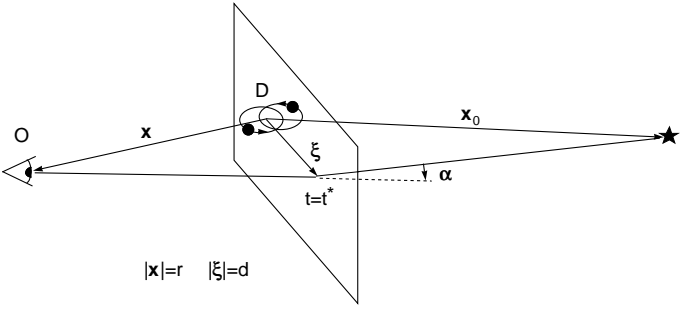}  \\
\caption{\label{fig1} The light from a source $S$ is lensed by a
  deflector ($D$). The impact parameter $\vec{\xi}$ is a vector in the
  lens plane. The distances of $D$ from the observer and from the
  light source respectively are $r$ and $r_0.$ }
\end{center}
\end{figure}

In ~\cite{kope,gwinn} the problem is faced in full generality,
determining the time delay, $\Delta$, and deflection angle,
$\vec{\alpha},$ caused by a localized source of gravitational
radiation, $\mathbf{D},$ acting as a deflector
of light (see Fig.~\ref{fig1}).  In~\cite{kope,gwinn}, a multipole
expansion for 
the energy momentum tensor  of the source is used. In~\cite{gwinn}
  this expansion is truncated at the quadrupole
and the resulting gravitational field is expressed in terms of the
deflector mass, its angular momentum and quadrupole.

In this paper we study the effect of the lens quadrupole not on
gravity wave production, but  on the scalar potential which is
responsible for lensing. Especially, we want to compute its
contribution to microlensing.
The quadrupole is the lowest multipole contributing to the emission of
gravitational waves. The scalar potential,
however, contains also the larger monopole and dipole
contributions. Na\"ively, one might expect that therefore the
quadrupole is irrelevant, unobservable. We shall show here that this
is not the case. For impact parameters close to the critical line,
which are those with large amplification, the quadrupole contribution
to the amplification is significantly enhanced and can become observable. 
Furthermore, mass and angular momentum are conserved quantities, while the
quadrupole is in general time dependent. It will therefore
introduce a time-dependence in the microlensing signal which we
determine.  A direct detection of such a modulation can
give important information about the  quadrupole of the lens and
therefore on the gravitational radiation it emits. This can be very
useful for the observation of gravitational radiation from the system.
But even if the effect is not observed directly, because of the
under sampling of data or because the period 
of the lens system is of the same order than the time-scale of
microlensing event, it has to be taken into account as a possible
source of error in static parameters estimations, especially for high
amplification events.
We conclude the paper with a rate estimation, taking into account
only compact binaries of our galaxy as example. However, we observe
that, more generally, such a modulation effect can be given by any
object with a varying quadrupole. Our result gives a new tool to
determine the quadrupole of a unknown system independent of its
nature. 

We stress that the modulation can be detected only if the period of the
system is smaller than the time-scale of microlensing {\em i.e.}
for relatively compact binaries with periods of less than about 30
days.

\section{The amplification factor}

We work in the thin lens approximation, which means that we may
project the lens mass distribution into a plane and we  consider
impact parameters $d = |\vec{\xi}|$ much smaller than the
distances $r$ and $r_0$ in Figure~1. Furthermore, we assume the
condition $\frac{\omega d^2}{c r}\ll1$ where $\omega$ is the
frequency of the binary, so that retardation inside the lens plane
can be neglected. We employ the center of mass system, {\em i.e.}
the coordinate system where the center of mass of the binary is at
rest at position $\vec{x}=0$. Up to the quadrupole, the
gravitational lens potential is then given by~\cite{gwinn}
\be\label{Psi} \Psi(\vec{\xi},t^*)=\left[
M+\epsilon_{jpq}k_pS_q\partial_j+
\frac{1}{2}I_{pq}^{TT}(t^*)\partial_p\partial_q\right]\ln d~, \ee
where we have set $G = c = 1$ and $t^*$ denotes retarded time,
$t^* = t-r$. Here $\vec{k}$ is the unit vector pointing from the
source to the observer, $M$ is the total mass of the system,
$\vec{S}$ is its angular momentum and $I_{pq}^{TT}$ is the
transverse traceless quadrupole tensor, projected into the plane
normal to $\vec{k}$,
\begin{eqnarray}
S^q(t) &=& \frac{1}{2} \epsilon^{qpr}\int d^3x
\left( x^pT^{0r}(\vec{x},t)- x^rT^{0p}(\vec{x},t)  \right) ~,\\
I_{pq} &=& \int d^3x \rho(\vec{x},t)(x_qx_p -\frac{1}{3}|x|^2\delta_{qp}),\\
I_{pq}^{TT} &=& \Big[\de_{ip}\de_{jq} +\frac{1}{2}(\de_{pq}+k_pk_q)k_ik_j
 \nonumber \\  &&
- (\de_{pi}k_qk_j+\de_{qi}k_pk_j)\Big]I_{ij}.
\end{eqnarray}
$T^{\mu\nu}(\vec{x})$ denotes the energy momentum tensor of the
source and $\rho(\vec{x},t) = T^{00}$ is the energy density. Since the
background spacetime is Minkowski, spatial index positions are
irrelevant. The time delay and the deflection angle can be expressed in
terms of $\Psi$ as~\cite{gwinn}
\be
\Delta=-4\Psi+2M\ln(4rr_0)~,\qquad\alpha_i=4\partial_i\Psi~.
\ee
The amplification of a far away light source with $r_0\gg r$, is now easily determined as
$\mu=\frac{1}{\det(A)}$ where $A$ is the Jacobian of the lens map (see
e.g.~\cite{Sc}),
 \be\label{Aij}
 A_{ij}=\delta_{ij}-4r\partial_i\partial_j\Psi(\vec{\xi},t^*).
 \ee

Without loss of generality, we fix the orientation of the
coordinate system so that the ${x_1}$ axis is aligned with the
impact vector and the third axis is normal to the lens plane.
Hence $\xi_1=d$, $\xi_2=0$ and $k_1=k_2=0$, $k_3=1$.
From Eqs.~(\ref{Psi},\ref{Aij}) one then obtains the following
expression for $\mu$,
\begin{eqnarray}\label{mu}
&&\mu^{-1} = \det(A)= 1-\frac{16r^2}{d^2}\left[ \frac{M^2}{d^2}
 + 4\frac{S_1^2+S_2^2}{d^4}+ \right.\nonumber\\
 &&+4M\left(\frac{S_2}{d^3} + \frac{3(I_{11}+
\frac{1}{2}I_{33})}{d^4}\right) +
24\frac{S_2(I_{11}+\frac{1}{2}I_{33})}{d^5}
 \nonumber\\
&& \left. -24\frac{S_1 I_{12}}{d^5}+
\frac{36(I_{11}+\frac{1}{2}I_{33})^2}{d^6} +
36\frac{I_{12}^2}{d^6} \right]~.
\end{eqnarray}
We use that in our coordinate system $I^{TT}_{ij}$ is entirely
determined by $I^{TT}_{11} = - I^{TT}_{22}= I_{11}+
\frac{1}{2}I_{33}$ and $I^{TT}_{12} = I_{12}$. The quadrupole
tensor $I_{ij}$ has to be evaluated at retarded time $t^*=t-r.$
T
he largest term containing the quadrupole $\propto  12(I_{11}+
\frac{1}{2}I_{33}/d^4$ is suppressed with respect to the monopole
$\propto M^2/d^2$, by a factor $\frac{|I_{11}|}{d^2M}$.
If $a$ denotes the mayor half axis of the binary, we have
$|I_{ij}|\simeq Ma^2$, hence the suppression is of the order of
$\ep^2 = (a/d)^2$. Therefore, one might suggest that systems with
large orbits have the strongest contribution from the quadrupole.
This is true, but in this case the time variation may not be
visible if the period of the system is larger than the duration of
the event, $T=2\pi a^{3/2}/\sqrt{M} \ge d/v$. Here $v$ is the
source velocity (projected into the lens plane).

Furthermore, our expansion breaks down at $a\simeq d$ since higher
multipoles can no longer be neglected and the microlensing
event probes the full matter distribution of the lens. This more
complex phenomenon has been studied extensively in the literature, see e.g.
Refs.~\cite{SchneiderWeiss} to \cite{uda}. Here we restrict ourselves
to $a/d<0.3$, say. We are mainly interested in compact binaries which
usually also generate a significant amount of gravitational waves.

The amplification is largest close to the critical line defined by
$\det(A)=0$. Neglecting the sub-dominant contributions this
corresponds to the Einstein radius $r_E = 2\sqrt{Mr} \equiv d_c$,
the critical impact parameter.  For large amplification, $\De =
|d-d_c|/d_c\ll 1$, the effect
of the quadrupole is enhanced by a factor
$\Delta^{-1}$. To illustrate this, we consider a
binary with angular momentum normal to the lens plane, and take
into account only the dominant contribution from the quadrupole in
Eq.~(\ref{mu}), $\frac{M}{d}\frac{12I^{TT}_{11}}{d^3}$. We parameterize
the quadrupole term by
$$
\frac{12(I_{11}+\frac{1}{2}I_{33})(t^*)}{d^3} =
\ga(t^*)\frac{M}{d}\ep^2~.
$$ 
Here $\ga(t^*)$ is a dimensionless function of order unity which
contains the information about the time dependence of the quadrupole
and hence, e.g. on the frequency of the gravitational gave emitted by
the system. The amplification can then be
approximated by \be\label{muapprox} \mu^{-1} \simeq 1-
\(\frac{d_c}{d}\)^4\[1+\ga\ep^2\]~, \quad \ep =\frac{a}{d} ~. \ee
We want to consider the case $d=d_c(1+\Delta)$ with $\Delta\ll 1$.
In this case we have to lowest order in the small parameters $\ep$
and $\Delta$ \be\label{muapprox2} \mu \simeq \frac{1}{4\Delta
-\ga\ep^2}  = \frac{\Delta^{-1}}{4 -\ga\ep^2/\Delta} ~. \ee We
conclude that the contribution from the quadrupole is significant
if the ratio $\ep^2/\Delta\simeq 4\mu\ep^2$ is significant, say
larger than a few percent.

As usual, our ray optical approach gives rise to a divergence of the
amplification when the impact parameter approaches the critical value
$d_{c} = 2\sqrt{M\cdot r}$.
The amplification grows indefinitely when $d$ decreases toward the
critical value. At distances smaller than $d_c$ there are in principle
multiple images, but they are too close together to be resolved by
present optical telescopes.
The divergences  in the geometrical optics treatment is
removed in the correct treatment  using wave optics~\cite{Sc}.

\section{Examples}

\begin{figure}
\begin{center}
\includegraphics[width=7cm,height=7cm]{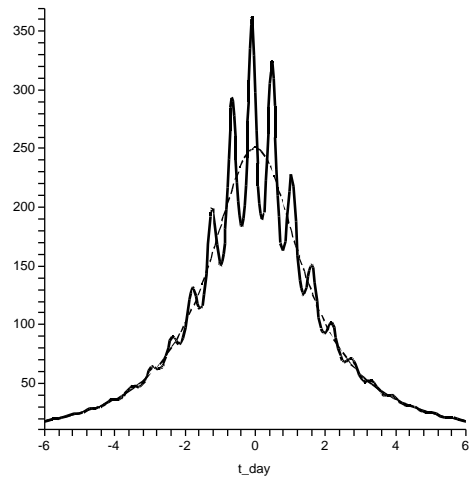}
\caption{ \label{fig2}The amplification is plotted as a function of
  time. The dashed
line represents the mass (static) contribution and the solid line is the total
(mass + quadrupole) signal. This corresponds to a microlensing event by a
neutron star or white dwarf binary with parameters $ T=10^5s=1.2$day and
  $r=200pc$. The impact parameter is $d=(1+10^{-3})d_c$.}
\end{center}
\end{figure}

\begin{figure}
\begin{center}
\includegraphics[width=7cm,height=7cm]{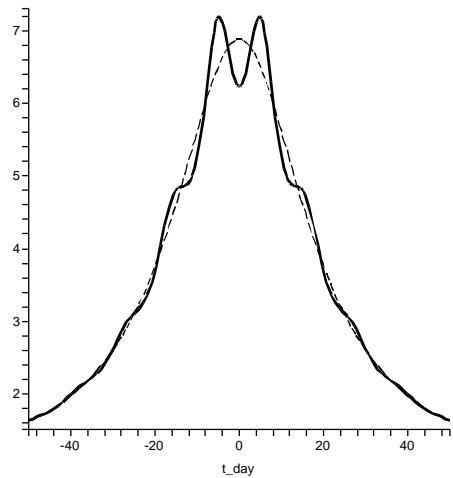}
\caption{\label{fig3} As Fig.~\ref{fig2}, but for a binary
with parameters $T=23$ days, $r=700pc$, $d=(1.04)d_c$.}
\end{center}
\end{figure}

In this section we present some examples for the modulation of the
magnification by microlensing by compact binaries in our galaxy. 
It is worth to stress that such a modulation can be produced by any
system having a varying quadrupole, and our result can be applied to
more general cases than the ones discussed here.
In Figs.~\ref{fig2} and~\ref{fig3} we choose for the BS two equal
masses $M_1=M_2= 1.4M_{\odot}$ in circular orbit in the lens 
plane so that the spin is aligned with the 3-axis, $S_1=S_2=0$. We consider a
background source moving with 100km/s relative to the lens $\bf D$. \\
In Fig.~\ref{fig2} the amplification is plotted as function of time
for a neutron star or white dwarf binary. A rotation period of
$T=10^5$sec, corresponding to an orbital radius
of $a\simeq 4.5\times10^{6}$km is assumed. The binary is placed
at distance $r=200$pc. The impact parameter is $d =(1 +
10^{-3})r_E$ yielding an amplification of about $\mu \simeq 250$. This
is very close to the critical impact parameter
$d_c = r_E = 2.2\times 10^8$km which gives infinite amplification
(within a ray optical treatment).
The quadrupole modulation amounts to $43\%$ of the static contribution at
maximum amplification. Our na\"\i ve estimate gives a relative
contribution $4\mu\ep^2 \simeq 0.4$ from the quadrupole, which is in
the right ballpark.\\
In the second example, plotted
in Fig.~\ref{fig3}, we take a deflector somewhat further away, $r=700$pc, and
consider a binary with period, $T=2\times 10^{6}$sec $=23$days. For
this system,
$a\simeq 3.4\times 10^{7}$km. The impact parameter of this case is
$d= 1.04r_E$ yielding a  maximal amplification of about 7. The
quadrupole modulation amounts still to
11$\%$ at maximum. Here the modulation signal is less significant since
$\Delta =(d-d_c)/d_c$ is larger, and also the period of the system is
significantly longer.

\begin{figure}
\begin{center}
\includegraphics[width=7cm,height=7cm]{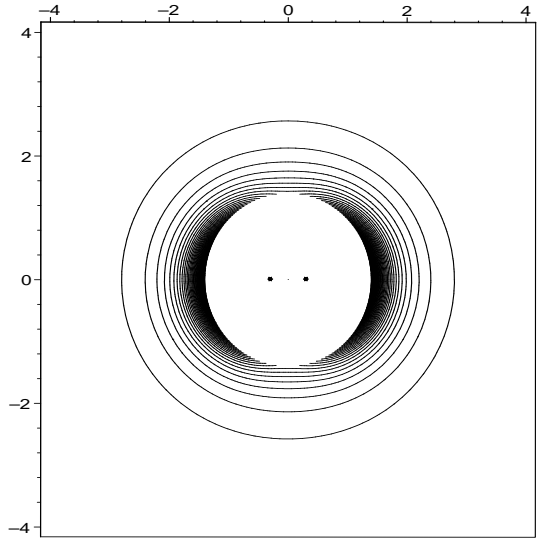}
\caption{\label{fig4} Example of lines of equal amplification for a
  binary system. The dots indicate the position of the stars and the
  center of mass respectively.
}
\end{center}
\end{figure}

In order to gain a good {\em qualitative} intuition of this phenomenon one
can visualize the lines of equal amplification for a lens system. For
a point-mass they are circles around 
the deflector. If a non-vanishing quadrupole moment is present, the
circle of divergence as well as the circles of equal amplification are
deformed as shown in Figure~\ref{fig4}. They are simply the solutions
of the equation $\mu^{-1} = 0$  and $\mu^{-1} =$ constant respectively,
as a function of the angle between $\vec\xi$
and the direction of the vector relating the two stars of the binary.
This vector rotates with the period of the system and with
it the deformed circles. The observed modulation of the amplification
comes from the rotation of the non-circular curves of constant amplification
in the lens plane. The deformation of the circles  is quite faint but
since the amplification becomes so large close to the critical line, 
the quadrupole is nevertheless observable. Furthermore, our
{\em quantitative} examples show that quadrupole modulations of high
importance can appear using quite reasonable parameters for the
lensing system, and for this reason, even if the modulations cannot be
detected directly, they have to be included in the error 
budget for microlensing events with high magnification, $\mu\ge
7$, if one wants to reach an accuracy in the predicted light curve
of about 10\% or better.
 
We stress that the modulation can only be observed
if the period  of the binary is shorter than the time interval over which the
amplification is significant.\\

It is interesting that observing such an event in the LISA range of
 frequencies, permits at the same time to determine
 the frequency and direction
 of the binary as source for gravitational waves. This will allow to detect the
 corresponding gravity wave signal out of the confusion noise in the
 LISA data~\cite{lisa}.

The idea to detect gravity waves via microlensing has been studied
in~\cite{lar}, and more recently in~\cite{rag},but in these works the gravity wave
source and the static
lens are two different objects, the first being far enough from  the second to
make the quadrupole time dependence discussed here unimportant.\\
In this paper we consider the quadrupole variation of the deflector
itself and we study
its contribution to the scalar lens potential. The effect from the also emitted
dynamical gravitational wave is much smaller than the one considered
here in the frequency
range we are interested in ($10^{-6}-10^{-3}Hz$), since it is
proportional to the second time derivative of the quadrupole. \\
However, measuring the modulation of $\mu$ provides access to information
about the variation of $I_{ij}$ itself. For example, it allows to
predict the frequency and amplitude of the gravity waves emitted by
the system.

As we have seen above, in order for the modulation to be measurable,
the microlensing event has to reach rather high magnification.

\begin{figure}[ht]
\begin{center}
\includegraphics[width=7cm]{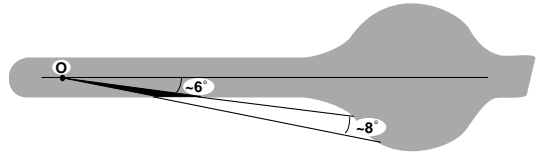}
\caption{\label{fig5} Configuration for the bulge survey. The black
  triangle represents the volume used to determine $N_b$.
}
\end{center}
\end{figure}

\section{Experimental forecast}

We focus on a galactic bulge survey where the observed
field is about $8^o\times 8^o =(\Delta\varphi)^2$ centered on galactic coordinates
$l\sim 4^o$, $b\sim -6^o$, which contains about $N_{s}=5\cdot10^7$ bulge
stars~\cite{Macho}. Using the binary population model of
Tutukov--Yungelson~\cite{tutu}
we can estimate the number of galactic black hole, neutron star and
white dwarf binaries to be about $3\times 10^8$.
Within the volume swept  by the light rays coming
from our sources, we expect to find about a fraction (see Fig.~\ref{fig5})
$$
 x\sim \frac{r^3(\Delta\varphi)^2}{3D_\mathrm{gal}R_\mathrm{gal}^2\pi}
\simeq 10^{-4}$$
of these
binaries, leading to $N_b \simeq 27000$. Here $R_\mathrm{gal}
\simeq 15$kpc is the radius of the galactic disk and $D_\mathrm{gal}
 \simeq 300$pc is its thickness and $r\sim
 D_\mathrm{gal}/2\frac{1}{\sin(6^o)} =1.4$kpc is the apparent
 thickness of the galactic disc in the direction of the survey.\\
The cross section $\sigma$ of the events for which the modulation
is visible is about $\sigma = 0.6d_c^2$. With this the fraction
$f$ of the observed field covered by a binary per unit time is
given $f= \sqrt{\sigma}v/A$, where $v$ is the center of mass
velocity of the source with respect to the binary and $A
=r^2(\Delta\varphi)^2$ is the area of the observed field. With
$r=700$pc and $v=100$km/sec we obtain an event rate \be
 \eta= \frac{\sqrt{\sigma}vN_{b}N_{s}}{A} \sim 0.1\ \rm{events/year} \ .
\ee
Note also, that we have taken into account only microlensing  by
compact binaries. Main sequence binaries which are sufficiently close,
so that $a/d<0.3$, say may very well contribute a more substantial event
rate. Furthermore, the modulations might be due to different objects
with a varying quadrupole, e.g cosmic strings~\cite{schild}.  

This event rate seems not out of reach of observations and it is already
 interesting to investigate whether such an event is not
 present in existing microlensing surveys, that is whether a
varying quadrupole behavior may fit one of the exotic microlensing
events detected so far.

Even if the modulation cannot be detected directly, the
contributions from the quadrupole and the angular momentum are
sufficiently important that they have to be included in the error
budget for microlensing events with high magnification, $\mu\ge
7$, if one wants to reach an accuracy in the predicted light curve
of about 10\%.

Let us finally estimate the contribution of the dipole.
The dominant effect coming from the spin is the term $4MS/d^3$.
Parameterizing $|S| \sim Ma^2\om \sim Mav_o$, where $\om$ denotes the
frequency of the binary and$v_o\sim \om a$ is the
orbital velocity. Hence $v_o \simeq
10^{-3}\sqrt{\frac{M}{3M_\odot}\frac{10^7\mathrm{km}}{a}}$
Its relative contribution is of the order of $MSd^{-3}/(M^2d^{-2})
\simeq (a/d)a\omega = \ep v_o \ll 1$, where $v_o$ denotes the
orbital velocity of the binary, 
$$
v_o \simeq \sqrt{\frac{M}{a}} \simeq 
10^{-3}\sqrt{\frac{M}{3M_\odot}\frac{10^7\mathrm{km}}{a}}~. 
$$
When
$d$ approaches $d_c$ also this term is parametrically enhanced
leading to a magnification 
\be\label{spin} \mu_S \simeq
\frac{\Delta^{-1}}{4(1+\ep v_o/\Delta)}~. \ee 
This term is
significant only for very compact and therefore very fast
binaries, $a \sim 10^3$km --- $10^4$km, which then have to be
sufficiently close so that $d_c$ is relatively small and $a/d$ is
still significant for $d\simeq d_c$. More precisely one finds 
$$
\ep v_o = \sqrt{\frac{R_sa}{2r}}\frac{d_c}{d} ~.
$$ 
Hence we need
$\frac{a}{2r}\ge 10^{-2}/\mu^2$ for the spin amplification to
amount to at least $10\%$. This looks quite unreasonable for
amplifications which are not gigantic. This term is also more difficult to
disentangle from the monopole since it is time independent like
the latter.

\section{Conclusion}

In this paper we have derived and discussed a new effect which leads to a
modulation of the light curve on microlensing events from compact
time dependent source where the impact parameter $d$ is larger than
the size of the source.
A simple analysis of the amplification modulation of such
an event allows to determine important parameters of the 
lens quadrupole such as its frequency and its amplitude. These can then be
used to predict the gravitational radiation emitted from the system.

This effect is of particular importance in microlensing events by 
compact binaries.  The relative contribution to the magnification
from the  quadrupole being of the
order of $4\mu(a/d)^2$, the effect is most significant for high
magnification. A rate estimation for galactic compact binaries shows
that typical microlensing surveys towards the galactic bulge should
detect about one such event every decade. But even in cases 
where it is not observed directly, the effect has to be included in the
error budget for the microlensing light curve.

 This work is supported by the Swiss National Science Foundation (FNS).


\end{document}